# Realization of an all-dielectric zero-index optical metamaterial

Parikshit Moitra[1†], Yuanmu Yang[1†], Zachary Anderson[2], Ivan I. Kravchenko[3], Dayrl P. Briggs[3], Jason Valentine[4*]

[1]Interdisciplinary Materials Science Program, Vanderbilt University, Nashville, Tennessee 37212, USA
[2]School for Science and Math at Vanderbilt, Nashville, TN 37232, USA
[3]Center for Nanophase Materials Sciences, Oak Ridge National Laboratory, Oak Ridge, Tennessee 37831, USA
[4]Department of Mechanical Engineering, Vanderbilt University, Nashville, Tennessee 37212, USA
[†]these authors contributed equally to this work
[*]email: jason.g.valentine@vanderbilt.edu

**Metamaterials offer unprecedented flexibility for manipulating the optical properties of matter, including the ability to access negative index[1–4], ultra-high index[5] and chiral optical properties[6–8]. Recently, metamaterials with near-zero refractive index have drawn much attention[9–13]. Light inside such materials experiences no spatial phase change and extremely large phase velocity, properties that can be applied for realizing directional emission[14–16], tunneling waveguides[17], large area single mode devices[18], and electromagnetic cloaks[19]. However, at optical frequencies previously demonstrated zero- or negative-refractive index metamaterials require the use of metallic inclusions, leading to large ohmic loss, a serious impediment to device applications[20,21]. Here, we experimentally demonstrate an impedance matched zero-index metamaterial at optical frequencies based on purely dielectric constituents. Formed from stacked silicon rod unit cells, the metamaterial possesses a nearly isotropic low-index response leading to angular selectivity of transmission and directive emission from quantum dots placed within the material.**

Over the past several years, most work aimed at achieving zero-index has been focused on epsilon-near-zero metamaterials (ENZs) which can be realized using diluted metals or metal waveguides operating below cut-off. These studies have included experimental demonstrations in the microwave[9,14], mid-IR[13], and visible regimes[12]. ENZ metamaterials have a permittivity ($\varepsilon$) that is near-zero and a permeability ($\mu$) of unity, resulting in a near-zero refractive index ($n = \sqrt{\mu\varepsilon}$). However, since the permeability remains finite, the relative optical impedance, ($Z = \sqrt{\mu/\varepsilon}$), is inevitably mismatched from free-space, resulting in large reflections from the interface. Impedance matched zero-index metamaterials (ZIMs) in which both the permittivity and permeability are set to zero eliminate these strong reflections and have recently been demonstrated at optical frequencies using metal based fishnet structures[11]. However, fishnet metamaterials do not possess isotropic optical properties and the use of metals inevitably introduces ohmic loss that will limit the thickness of the material.

Resonant all-dielectric metamaterials offer one potential solution to both of these issues[22–24]. Formed from high refractive index resonators, dielectric metamaterial unit cells support an electric and magnetic dipole response due to Mie resonances. Proper control of the lattice arrangement, resonator geometry, and composition allows control over the effective permittivity and permeability of the metamaterial. Due to the absence of ohmic loss, dielectric metamaterials can be much less absorptive than their metallic counterparts and their simple unit cell geometries offer the possibility to achieve 3D isotropic metamaterials[25], a task which has proved challenging utilizing more complicated metal-based unit cells[21]. However, while magnetic modes in high-index particles have recently been experimentally characterized at optical frequencies[26,27], implementations of dielectric metamaterials have so far been limited to the microwave[25,28,29] and mid-infrared regimes[30].

Here, we report the first experimental demonstration of an all-dielectric ZIM operating at infrared frequencies. The metamaterial's design is based off of a recent proposal by Chan[10] in which it was shown that a metamaterial made of purely dielectric high-index rods can exhibit a Dirac cone at the Γ point in the band structure, a feature that is similar to the electronic band structure in graphene[31]. At the Dirac point, the metamaterial exhibits zero effective permittivity and permeability, resulting in an impedance matched ZIM. Experiments at microwave frequencies demonstrated some of the unique properties arising from an effective index of zero[10], such as cloaking and lensing, however, a demonstration at optical frequencies has remain unexplored. Here, we implement the ZIM at optical frequencies using vertically stacked silicon (Si) rods, allowing access from free-space. We demonstrate that the optical ZIM serves as an angular optical filter while also enhancing spontaneous emission, both in directivity and rate, from quantum dot light sources embedded inside the structure. The experimental results, along with numerical calculations, serve as direct evidence of impedance matched near-zero-index within the metamaterial.

The fabricated ZIM consists of 200 μm long Si rods which support the magnetic and electric dipole resonances and the rods are separated by a low index material, silicon dioxide (SiO$_2$) (Fig. 1a). Fabrication began with a multilayer stack of 11 alternating layers of α-Si ($\varepsilon$=13.7, 260 nm thick) and SiO$_2$ ($\varepsilon$=2.25, 340 nm thick) followed by a patterning and reactive-ion etching (RIE). In the final step, poly(methyl methacrylate) (PMMA) ($\varepsilon$=2.23) was spin-coated onto the sample to fill the air gaps. Figure 1b shows a cross-section of the fabricated structure before final PMMA spin-coating and the inset depicts a sample after spin-coating. A total of 5 functional layers (Si / SiO$_2$ pairs) results in a metamaterial with a thickness of 3 μm, about twice the free-space wavelength at the zero-index point.

The band structure corresponding to the bulk ZIM (infinitely thick), consisting of a stack of square cross-section Si rods embedded in SiO2 with $w = t$ = 260 nm and $a$ = 600 nm, is shown in Fig. 2a. The band structure is computed for TM polarization with the electric field oriented along the rod axis and the Dirac cone like dispersion can be observed at the center of Brillouin zone where two transverse bands with linear dispersion intersect a flat longitudinal band, resulting in triple degeneracy. We utilized field-averaging of the Bloch modes[32] to retrieve the metamaterial's effective bulk optical properties (Fig. 2b) and simultaneous zero-permittivity and permeability are obtained at the triple degeneracy frequency of 211 THz ($\lambda_0$=1422 nm). This response is directly attributable to the strong magnetic and electric dipole Mie resonances within the Si rods (inset of Fig. 2b). In addition, a relatively broadband impedance matched low-index region is present around the zero-index point. Over the frequency range from 215 THz to 225 THz there exists only one propagating band, TM4, which allows us to define an effective index of refraction for the structure. The isofrequency contours (IFCs) over this frequency range are shown in Fig. 2c and it can be observed that they maintain a nearly circular shape indicating that the material maintains a relatively isotropic low-index response over this bandwidth. This nearly isotropic response is critical for preserving the unique physics associated with low- or zero-index materials such as directionally selective transmission and emission from within the material.

While these bulk parameters are used as a design guide, the fabricated metamaterial has both a finite thickness as well as non-uniform rod sizes and will thus deviate somewhat from the bulk parameters. To better understand the optical response of the fabricated metamaterial, full-wave finite-difference time-domain (FDTD) simulations were performed and S-parameter retrieval[33,34] was used to compute the effective optical properties (Fig. 3a,b). Due to the non-uniform rod size, the impedance matched point in fabricated structure is shifted to 1400 nm corresponding to $n_{eff} = \mu_{eff} = \varepsilon_{eff} = 0.12$. It can also be observed that a metallic region is opened from 1430 nm to 1460 nm in which the permittivity is negative while the permeability remains positive. The metamaterial exhibits a negative index region beginning at 1460 nm though this region cannot be assigned with isotropic optical properties due to presence of the longitudinal band which is accessible at large incident angles.

In order to gauge the agreement with the simulated material properties, transmittance of the fabricated ZIM was acquired by illuminating the sample with normal-incident white light with the electric field oriented along the Si rod axis (Fig. 3c). The simulated and measured transmittance show excellent agreement in spectral shape, though the experimentally obtained curve has a lower amplitude that is likely due to non-uniformity in rod width at the edges of the sample as well as surface roughness. The measured spectrum shows a peak transmission of 80% at 1405 nm, the spectral position corresponding to the impedance-matched low-index point. A dip in transmission also occurs at 1460 nm, corresponding to the metallic region of the sample.

One of the most fascinating properties of isotropic low-index materials is that light incident from free-space is only transmitted over a narrow range of incidence angles. This effect is a direct consequence of phase matching at the interface which requires that the wavevector along the interface be conserved. As illustrated with the IFCs depicted in Fig. 4a, in a low-index metamaterial, the wavevector is restricted to extremely small values causing light incident at high angles ($k_{y,0} > |k_{ZIM}|$) to be reflected while near-normal incident light is transmitted ($k_{y,0} < |k_{ZIM}|$). This effect is evident when examining the simulated transmittance with regard to wavelength and angle of incidence (Fig. 4b) for the fabricated ZIM. The material exhibits near-zero transmission at off-normal incident angle within the low positive index band centered at 1400 nm, with improving angular confinement in transmission as we approach the zero-index point. The presence of the longitudinal band within the negative index region ($\lambda_0 \sim 1475$ nm) is also apparent, allowing transmission at large incident angles.

To experimentally verify that the fabricated ZIM preserves these features, light transmitted through the sample was imaged in the Fourier plane. The illumination was focused with a large numerical aperture objective (NA=0.85), providing incident angles up to 58.2°. The acquired Fourier images (Figs 4. c-g) show confinement of light along the *y*-axis of the material which is in the direction of in-plane periodicity. Within the low index region between 1400 nm and 1475 nm, tighter confinement of $k_y$ is observed with progressive lowering of the refractive index, in agreement with the simulated data. Confinement is absent in the *x*-direction due to the fact that the electric field is no longer directly along the rod axis for these incident angles. The directional filtering of transmission demonstrates that a nearly isotropic low-index is indeed preserved within the fabricated metamaterial.

An additional consequence of the near-zero spatial phase change is that incoherent isotropic emitters placed within the ZIM will tend to emit coherently in the direction normal to the air/ZIM interface[14–16]. To demonstrate this effect, we first simulated the emission from a line source placed both in the center of the material as well as distributed throughout the volume of the material, emitting at a free-space wavelength of $\lambda_0$=1425 nm (Fig. 5b.) Both emission profiles are confined to an overall angular spread of 10° and are not sensitive to the position of the emitter. The two small side lobes at -60° and 60° are due to scattering from the corners of the metamaterial, an effect associated with having a material with finite area. For experimental demonstrations, we placed lead sulfide (PbS) semiconductor quantum dots (QDs) within the ZIM to act as the emitter. The QDs had a luminescence peak centered at 1420 nm and a full width at half maximum of 172 nm (see Supplementary Information for emission spectrum). The QDs were sandwiched between two PMMA layers within the ZIM, and were excited with a tightly focused 1064 nm laser beam. The Fourier-plane images (Fig. 5c-f) from both an unstructured PMMA/QD film and QDs placed within the metamaterial show good angular confinement in the *y*-direction and an over two-fold increase in intensity from the ZIM compared to the unstructured case. The increase in emission is a result of the uniform phase distribution within the ZIM, leading to constructive interference from emitters throughout the material[12], further supporting the realization of a near-zero refractive index. We also note that although the luminescence peak of the QDs matches the low-index band of fabricated ZIM structure quite closely, parts of the emission fall beyond the low-index band, resulting in slightly lower angular confinement compared with the transmission data.

Here we have experimentally demonstrated the first all-dielectric zero-index metamaterial at optical frequencies, which exhibits nearly isotropic low index response for a particular polarization resulting in angular selectivity of transmission and spontaneous emission. The realization of impedance matched ZIMs at optical frequencies opens new avenues towards the development of angularly selective optical filters, directional light sources, and large area single mode photonic devices. Furthermore, the advent of all-dielectric optical metamaterials provides a new route to developing novel optical metamaterials with both low absorption loss and isotropic optical properties.

**Methods**

**Simulations**

Numerical simulations were carried out with the measured permittivity value of single layer α-Si deposited on quartz. The band structure was calculated using MIT Photonic Bands and the fields at the boundary of the unit cell were extracted for computing the effective optical properties via field averaging, preserving the continuity of the tangential and normal averaged-field components. FDTD simulations (CST Microwave Studio) were used to calculate the complex transmission and reflection coefficients which were used in S-parameter retrieval. The transmittance with respect to wavelength and angle of incidence was computed with finite element method simulations (Comsol Multiphysics).

**Sample Fabrication**

5 functional layers of α-Si (260 nm) and $SiO_2$ (340 nm) with an additional $SiO_2$ layer of 170 nm (total 11 layers) were deposited on 4 inch quartz wafers using low pressure chemical vapor deposition. An etch mask composed of chromium was defined using electron beam lithography and RIE (Oxford Plasma Lab) was used to structure the multilayer film. During RIE the chemistry was adjusted with etch depth so that the alternating Si and $SiO_2$ layers could be patterned in a single process step. To ensure that PMMA completely filled the structure, a vacuum was used remove air bubbles followed by soft baking. Complete filling of the PMMA was confirmed via focused ion beam cross-sectioning at multiple points in the sample.

# References

1. Shelby, R. A., Smith, D. R. & Schultz, S. Experimental verification of a negative index of refraction. *Science* **292**, 77–79 (2001).

2. Zhang, S. *et al.* Experimental Demonstration of Near-Infrared Negative-Index Metamaterials. *Physical Review Letters* **95**, 137404 (2005).

3. Shalaev, V. M. *et al.* Negative index of refraction in optical metamaterials. *Optics Letters* **30**, 3356–3358 (2005).

4. Valentine, J. *et al.* Three-dimensional optical metamaterial with a negative refractive index. *Nature* **455**, 376–379 (2008).

5. Choi, M. *et al.* A terahertz metamaterial with unnaturally high refractive index. *Nature* **470**, 369–373 (2011).

6. Decker, M., Klein, M. W., Wegener, M. & Linden, S. Circular dichroism of planar chiral magnetic metamaterials. *Optics Letters* **32**, 856–858 (2007).

7. Plum, E., Fedotov, V. A., Schwanecke, A. S., Zheludev, N. I. & Chen, Y. Giant optical gyrotropy due to electromagnetic coupling. *Applied Physics Letters* **90**, 223113 (2007).

8. Zhang, S. *et al.* Negative Refractive Index in Chiral Metamaterials. *Physical Review Letters* **102**, 023901 (2009).

9. Liu, R. *et al.* Experimental Demonstration of Electromagnetic Tunneling Through an Epsilon-Near-Zero Metamaterial at Microwave Frequencies. *Physical Review Letters* **100**, 023903 (2008).

10. Huang, X., Lai, Y., Hang, Z. H., Zheng, H. & Chan, C. T. Dirac cones induced by accidental degeneracy in photonic crystals and zero-refractive-index materials. *Nature Materials* **10**, 582–586 (2011).

11. Yun, S. *et al.* Low-Loss Impedance-Matched Optical Metamaterials with Zero-Phase Delay. *ACS Nano* **6**, 4475–4482 (2012).

12. Vesseur, E., Coenen, T., Caglayan, H., Engheta, N. & Polman, A. Experimental Verification of n=0 Structures for Visible Light. *Physical Review Letters* **110**, 013902 (2013).

13. Adams, D. *et al.* Funneling Light through a Subwavelength Aperture with Epsilon-Near-Zero Materials. *Physical Review Letters* **107**, 133901 (2011).

14. Enoch, S., Tayeb, G., Sabouroux, P., Guérin, N. & Vincent, P. A Metamaterial for Directive Emission. *Physical Review Letters* **89**, 213902 (2002).

15. Ziolkowski, R. Propagation in and scattering from a matched metamaterial having a zero index of refraction. *Physical Review E* **70**, 046608 (2004).

16. Alù, A., Silveirinha, M., Salandrino, A. & Engheta, N. Epsilon-near-zero metamaterials and electromagnetic sources: Tailoring the radiation phase pattern. *Physical Review B* **75**, 155410 (2007).

17. Silveirinha, M. & Engheta, N. Tunneling of Electromagnetic Energy through Subwavelength Channels and Bends using ε-Near-Zero Materials. *Physical Review Letters* **97**, 157403 (2006).

18. Bravo-Abad, J., Joannopoulos, J. D. & Soljacic, M. Enabling single-mode behavior over large areas with photonic Dirac cones. *Proceedings of the National Academy of Sciences of the United States of America* **109**, 9761–9765 (2012).

19. Hao, J., Yan, W. & Qiu, M. Super-reflection and cloaking based on zero index metamaterial. *Applied Physics Letters* **96**, 101109 (2010).

20. Boltasseva, A. & Atwater, H. A. Low-Loss Plasmonic Metamaterials. *Science* **331**, 290–291 (2011).

21. Soukoulis, C. M. & Wegener, M. Past achievements and future challenges in the development of three-dimensional photonic metamaterials. *Nature Photonics* **5**, 523–530 (2011).

22. Lewin, L. The electrical constants of a material loaded with spherical particles. *Proc. Inst. Elec. Eng., Part 3* **94**, 65–68 (1947).

23. O'Brien, S. & Pendry, J. B. Photonic band-gap effects and magnetic activity in dielectric composites. *Journal of Physics: Condensed Matter* **14**, 4035–4044 (2002).

24. Zhao, Q., Zhou, J., Zhang, F. & Lippens, D. Mie resonance-based dielectric metamaterials. *Materials Today* **12**, 60–69 (2009).

25. Zhao, Q. *et al.* Experimental Demonstration of Isotropic Negative Permeability in a Three-Dimensional Dielectric Composite. *Physical Review Letters* **101**, 027402 (2008).

26. Kuznetsov, A. I., Miroshnichenko, A. E., Fu, Y. H., Zhang, J. & Luk'yanchuk, B. Magnetic light. *Scientific reports* **2**, 492 (2012).

27. Shi, L., Tuzer, T. U., Fenollosa, R. & Meseguer, F. A New Dielectric Metamaterial Building Block with a Strong Magnetic Response in the Sub-1.5-Micrometer Region: Silicon Colloid Nanocavities. *Advanced materials* **24**, 5934–5938 (2012).

28. Peng, L. *et al.* Experimental Observation of Left-Handed Behavior in an Array of Standard Dielectric Resonators. *Physical Review Letters* **98**, 157403 (2007).

29. Popa, B.-I. & Cummer, S. Compact Dielectric Particles as a Building Block for Low-Loss Magnetic Metamaterials. *Physical Review Letters* **100**, 207401 (2008).

30. Ginn, J. *et al.* Realizing Optical Magnetism from Dielectric Metamaterials. *Physical Review Letters* **108**, 097402 (2012).

31. Novoselov, K. S. *et al.* Two-dimensional gas of massless Dirac fermions in graphene. *Nature* **438**, 197–200 (2005).

32. Tsukerman, I. Effective parameters of metamaterials: a rigorous homogenization theory via Whitney interpolation. *Journal of the Optical Society of America B* **28**, 577–586 (2011).

33. Chen, X., Grzegorczyk, T., Wu, B.-I., Pacheco, J. & Kong, J. Robust method to retrieve the constitutive effective parameters of metamaterials. *Physical Review E* **70**, (2004).

34. Smith, D. R., Vier, D. C., Koschny, T. & Soukoulis, C. M. Electromagnetic parameter retrieval from inhomogeneous metamaterials. *Physical Review E* **71**, 036617 (2005).

**Acknowledgements**

This work was funded by the Office of Naval Research (ONR) under programs N00014-11-1-0521 and N00014-12-1-0571 and the United States-Israel Binational Science Foundation under program 2010460. A portion of this research was conducted at the Center for Nanophase Materials Sciences, which is sponsored at Oak Ridge National Laboratory by the Scientific User Facilities Division, Office of Basic Energy Sciences, U.S. Department of Energy. The authors thank Dr. Nick Lavrik for helpful discussions regarding RIE processing.

**Figures**

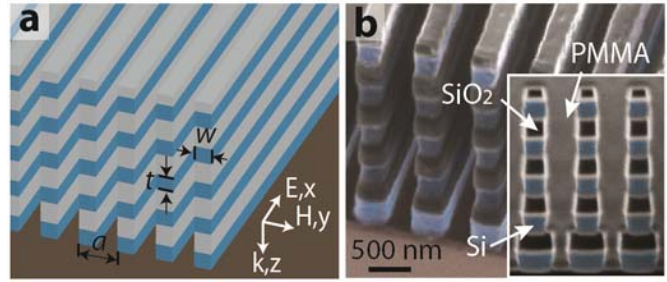

**Figure 1 Diagram and images of fabricated ZIM structure. a,** Diagram of the ZIM structure with a unit cell period of $a$ = 600 nm and $w = t$ = 260 nm. **b,** False color focused ion beam image of the ZIM before spin-coating PMMA. The inset shows a cross-section of the structure after PMMA filling. The fabricated sample has 11 alternating Si/SiO$_2$ layers with Si rod widths of 270 nm, 280 nm, 310 nm, 320 nm and 380 nm, from top to bottom.

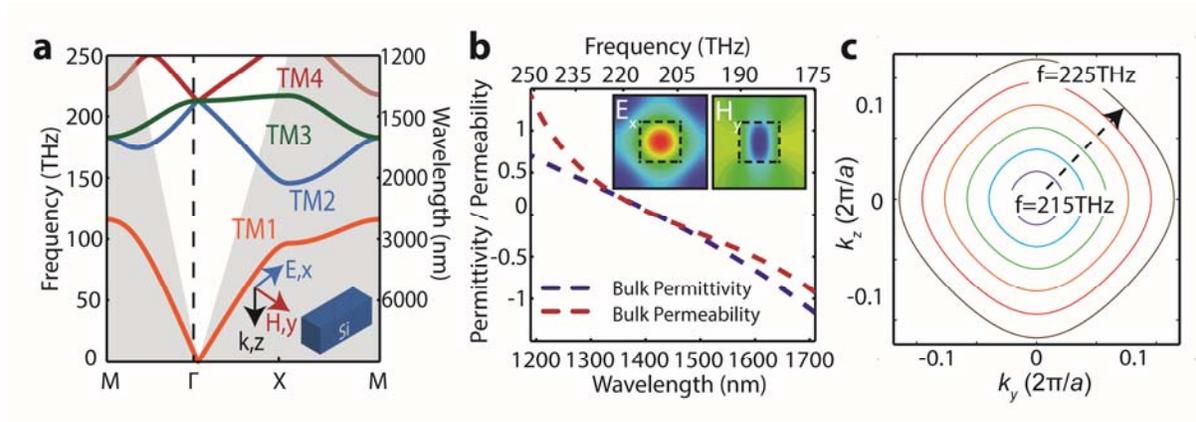

**Figure 2 Optical properties of the bulk ZIM. a,** Band diagram of uniform bulk ZIM (infinitely thick) for TM polarization. Dirac cone dispersion is observed at the Γ point with triple degeneracy at 211 THz. The shaded area denotes regions outside the free-space light line. **b,** Retrieved effective permittivity and permeability of the bulk ZIM acquired using field-averaging. The inset shows the electric and magnetic fields within a single unit cell at the zero-index frequency indicating a strong electric monopole and magnetic dipole response. **c,** IFC of the TM4 band. The contours are nearly circular (i.e., isotropic) for a broad frequency range and increase in size away from the zero-index frequency indicating a progressively larger refractive index.

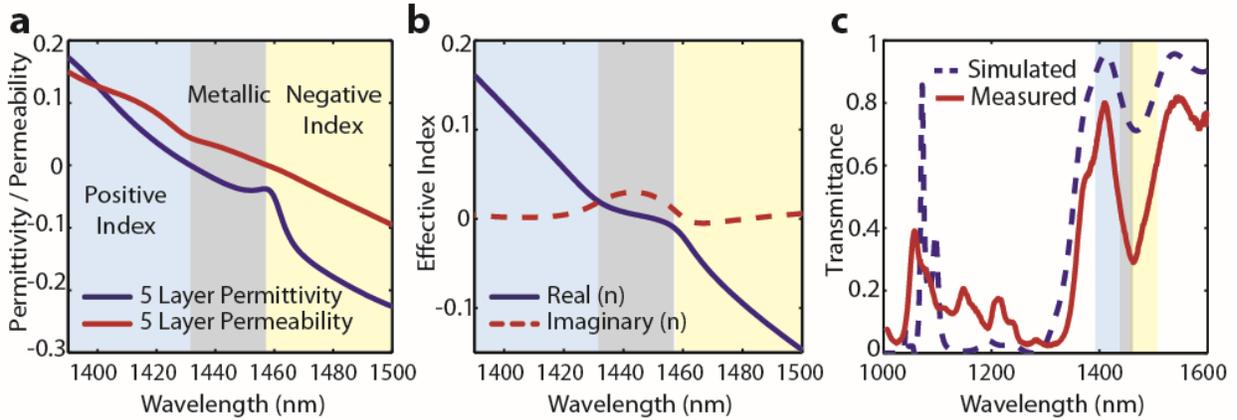

**Figure 3 Optical properties and transmittance of fabricated ZIM. a,** Effective permittivity and permeability of the fabricated ZIM obtained using S-parameter retrieval. Regions corresponding positive index, metallic properties, and negative index are denoted with blue, grey, and yellow shading, respectively. **b,** Effective refractive index of the fabricated structured obtained using S-parameter retrieval. **c,** Experimental (red) and theoretical transmittance (dotted blue) curves of the ZIM (200 x 200 μm$^2$ total pattern area).

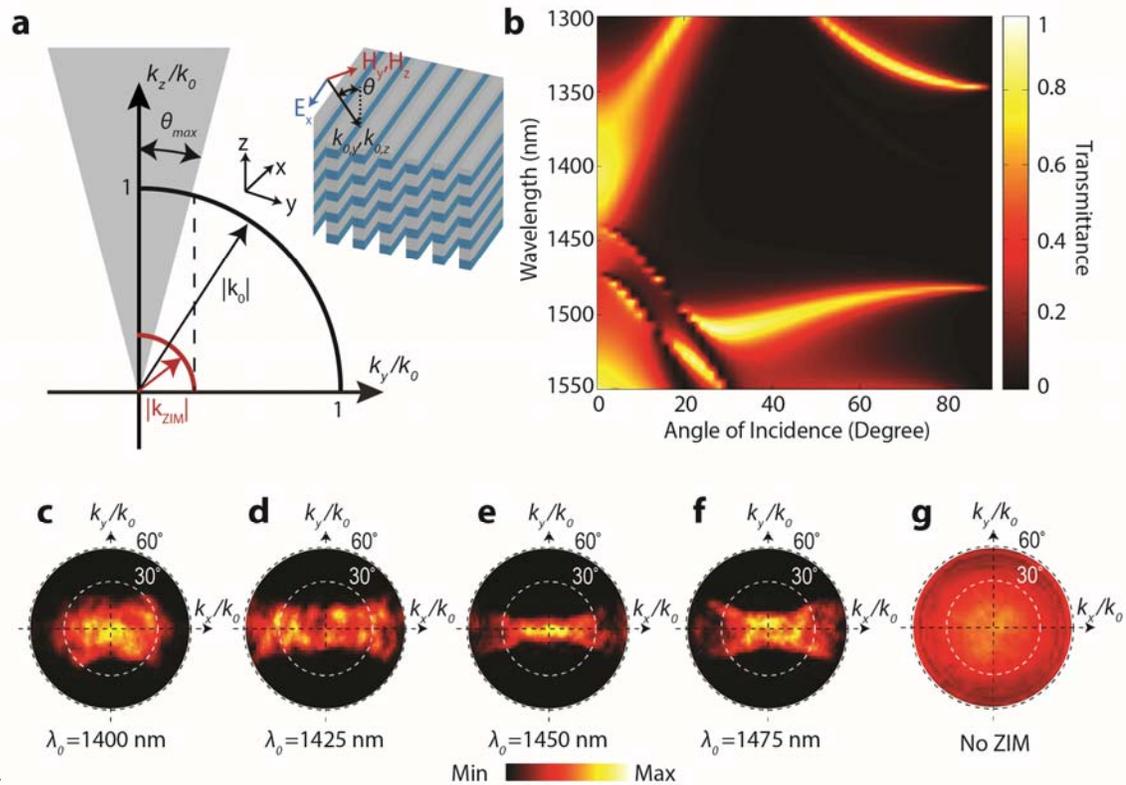

**Figure 4 Angular selectivity of transmission. a,** IFCs of air and a low-index metamaterial illustrating angularly selective transmission due to conservation of the wave vector parallel to the surface. **b,** Simulated angle and wavelength-dependent transmittance of the fabricated structure. **c-f,** Fourier-plane images of a beam passing through the fabricated ZIM structure within the low index band. Angularly selective transmission can be observed in the *y*-direction due to the low effective index. Along the *x*-direction, angular selectivity is not preserved due to the one-dimensional nature of Si rods. **g,** Fourier-plane image of the illumination beam demonstrating uniform intensity over the angular range measured.

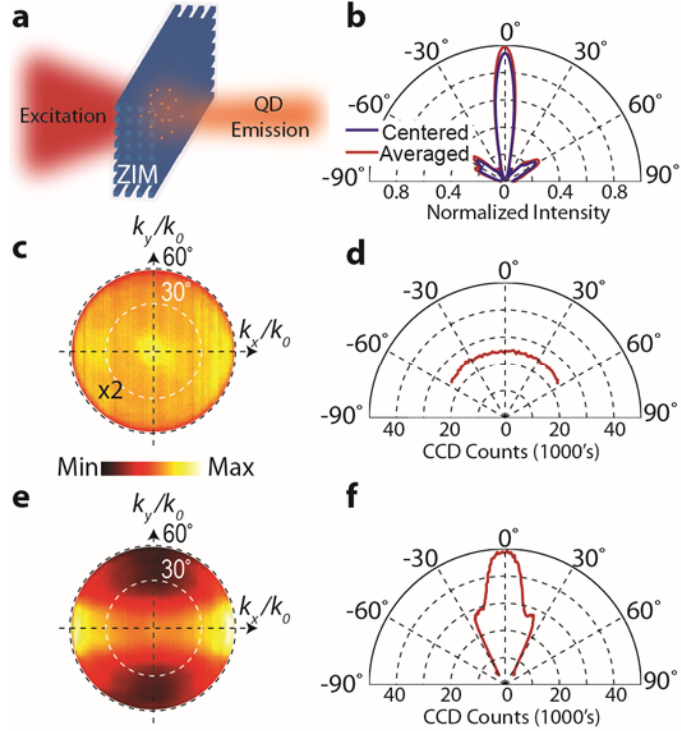

**Figure 5 Directional QD emission from within the ZIM. a,** Schematic of laser-pumped QD emission from within the ZIM structure. **b,** Calculated emission profile for a line source placed in the center of the material (centered) and the average profile from line sources placed throughout the material (averaged). **c,** 2D Fourier-plane images of quantum dot emission on the substrate, intensity is scaled by two times. **d,** A cross-section of the emission taken at $k_x = 0$. **e,** 2D Fourier-plane images of QD emission within the ZIM, respectively, showing enhanced rate and directivity of spontaneous emission. **f,** A cross-section of the emission taken at $k_x = 0$.